\documentclass[prb,showpacs,byrevtex,twocolumn,floatfix]{revtex4}
\usepackage{graphicx}
\usepackage{dcolumn}
\usepackage{bm}

\begin{document}

\bibliographystyle{apsrev}

\preprint{Draft copy --- not for distribution}
%
%
\title{Infrared optical properties of \boldmath Pr$_2$CuO$_4$ \unboldmath}
\author{C. C. Homes}%
\email[Electronic address: ]{homes@bnl.gov}%
\affiliation{Department of Physics, Brookhaven National Laboratory, Upton, NY
11973}
%
%
\author{Q. Li}%
\affiliation{Department of Materials Science, Brookhaven National Laboratory,
Upton, NY 11973}
%
%
\author{P. Fournier}%
\altaffiliation{Present address: D\'{e}partment de Physique,
 Universit\'{e} de Sherbrooke, Sherbrooke, Qu\'{e}bec, J1K 2R1 Canada.}
%
%
\author{R. L. Greene}
\affiliation{Center for Superconductivity Research, Department of Physics and
Astronomy, University of Maryland, College Park, MD 20742}%
\date{\today}
%
%
\begin{abstract}
The {\it ab}-plane reflectance of a Pr$_2$CuO$_4$ single crystal has been
measured over a wide frequency range at a variety of temperatures, and the
optical properties determined from a Kramers-Kronig analysis.
Above $\approx 250$~K, the low frequency conductivity increases quickly with
temperature; $\rho_{dc}\approx 1/\sigma_1(\omega\rightarrow 0)$ follows the
form $\rho_{dc}\propto \exp(E_a/k_BT)$, where $E_a\approx 0.17$~eV is much less
than the inferred optical gap of $\approx 1.2$~eV.  Transport measurements show
that at low temperature the resistivity deviates from activated behavior and
follows the form $\rho_{dc}\propto \exp [(T_0/T)^{1/4}]$, indicating that the
{\it dc} transport in this material is due to variable-range hopping between
localized states in the gap.
The four infrared-active $E_u$ modes dominate the infrared optical properties.
Below $\approx 200$~K, a striking new feature appears near the low-frequency
$E_u$ mode, and there is additional new fine structure at high frequency. A
normal coordinate analysis has been performed and the detailed nature of the
zone-center vibrations determined.  Only the low-frequency $E_u$ mode has a
significant Pr-Cu interaction. Several possible mechanisms related to the
antiferromagnetism in this material are proposed to explain the sudden
appearance of this and other new spectral features at low temperature.
\end{abstract}
\maketitle

%
%

%
%
\section{INTRODUCTION}
Within the family of high-temperature cuprate superconductors, the Ce-doped
series $R_{2-x}$Ce$_x$CuO$_{4-\delta}$, where $R$=Nd, Sm, Eu, Gd, etc., are the
only materials which appear to be electron doped.\cite{tokura89} The undoped
Nd$_2$CuO$_4$ and Pr$_2$CuO$_4$ are antiferromagnetic insulators, which become
``bad metals'' with Ce doping until the sudden onset of superconductivity at
$x\approx 0.14$.  The region of superconductivity in the electron-doped
materials is quite narrow ($x\approx 0.14-0.18$).\cite{takagi89a,takagi89b}
There is essentially no ``underdoped'' region for the superconductivity, with
the maximum value for the $T_c$'s of $23\,$K and $19\,$K in the Nd and Pr
systems, respectively, occurring at $x\approx 0.15$.  At higher dopings $T_c$
decreases rapidly vanishing above $x\approx 0.18$.  At the solubility limit
($x=0.22$) the Nd system is metallic with no evidence of
superconductivity.\cite{hagen91} The optimally doped systems become
superconducting only after oxygen reduction\cite{izumi89,klamut93,kawashima94}
($\delta\approx 0.01-0.03)$, and some transport measurements suggest that both
electrons and holes participate in the charge transport in the superconducting
phase.\cite{jiang94,fournier97} The role played by oxygen in these materials
may be more complex than in the hole-doped cuprates.   It is the interesting
behavior of these superconducting systems that motivates an examination of the
optical properties of one of the parent compounds, Pr$_2$CuO$_4$.

\begin{figure}
\vspace*{-0.7in}%
\centerline{\includegraphics[width=2.7in]{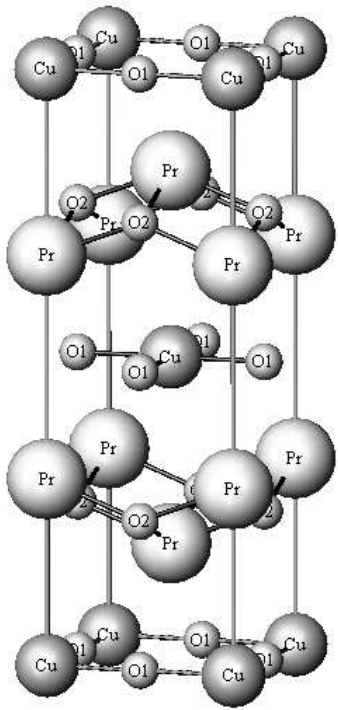}}%
\vspace*{-1.0in}%
\caption{The unit cell of the Pr$_2$CuO$_4$ in the $I4/mmm$ space group showing 
the square planar coordination of the Cu-O(1) layers, as well as the corrugated 
structure of the Pr-O(2) layers.  The unit cell dimensions are $a=b=3.943$~\AA\ 
and $c=12.15$~\AA\ (Ref.~\onlinecite{sumarlin95}).}%
\label{unitcell} 
\end{figure}

The $T^\prime$ structure of Pr$_2$CuO$_4$ is similar to the $T$ structure of
the hole-doped La$_{2-x}$Sr$_x$CuO$_4$; both structures are body-centered
tetragonal, space group $I4/mmm$ $(D_{4h}^{17})$ (Ref.~\onlinecite{muller75}).
These materials consist of two-dimensional sheets of copper-oxygen layers,
which define the {\it a-b} planes, with the {\it c} axis being perpendicular to
the planes. The $T$ and $T^\prime$ structures differ in the location of the
oxygen atoms between the copper-oxygen sheets.
In the $T$ structure the copper atoms have octahedral coordination, surrounded
by four oxygen atoms in the {\it a-b} plane, and two apical oxygens along the
{\it c} axis.
However, in the $T^\prime$ structure shown in Fig.~\ref{unitcell}, the apical
sites are empty and the out-of-plane oxygen atoms are not chemically bonded to
the copper atoms in the planes, which, as a result, have a square
coordination.\cite{crawford90a}  While the difference between the $T$ and
$T^\prime$ structures results in different Raman active modes, the same number
of infrared active modes $3A_{2u}+B_{2u}+4E_u$ are expected for
each.\cite{crawford90a} (The doubly degenerate $E_u$ modes are active in the
{\it a-b} planes, the singly degenerate $A_{2u}$ modes are active only along
the {\it c} axis, and the $B_{2u}$ mode is silent.)

%
%

%
%
The strong Cu-O bonding in the {\it a-b} plane of this material gives rise to
two-dimensional electronic and magnetic behavior. The weak out-of-plane
coupling induces long range antiferromagnetic (AFM) order in the Cu spins at
the relatively modest temperature of $T_{N,Cu}\approx 250-280$~K (Refs.
\onlinecite{cox89,matsuda90,sumarlin95}), which is similar to the values of
$T_{N,Cu}\approx 250-300$~K observed in Nd$_2$CuO$_4$
(Refs.~\onlinecite{matsuda90,endoh89}), Eu$_2$CuO$_4$
(Ref.~\onlinecite{chattopadhyay94}), and La$_2$CuO$_4$
(Refs.~\onlinecite{sun91,hayden91}).
%
%
The rare earth Pr ions carry localized $4f$ moments, which typically order at
very low temperature. However, from symmetry considerations, the Pr moments
should order below $T_{N,Cu}$ in this material.\cite{hill95} In fact, due to
the significant exchange interactions between the Pr ions which are mediated
through the copper-oxygen layers, there is a Pr contribution to the
susceptibility\cite{sumarlin95} below $\approx 200$~K.

The optical properties of Nd$_2$CuO$_4$ have been investigated in ceramics and
single crystals.\cite{degiorgi89,heyen90,zhang91,lupi92,jandl95,jandl96,
strach97,calvani96} Single crystals of Pr$_2$CuO$_4$ have been studied at
either room temperature \cite{cooper90,tajima91,arima93,jandl97} or at low
temperature\cite{crawford90a,crawford90b} ($\approx 10$~K), but there has been
no detailed investigation of the temperature dependence of the optical
properties of this material.  Furthermore, there is some disagreement in the
literature with respect to the vibrational parameters of the $E_u$ modes.

%
%
\begin{figure}
\centerline{\includegraphics[width=4.0in]{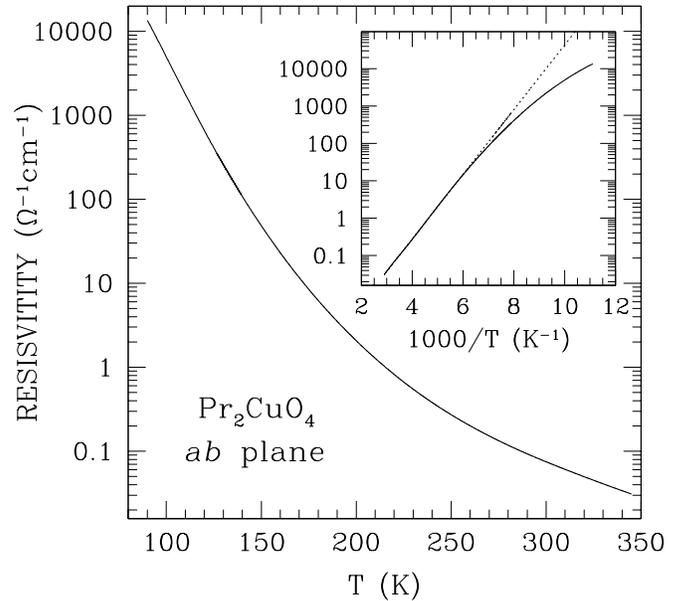}}%
\vspace*{-0.5in}%
\caption{The temperature dependence of the {\it ab}-plane resistivity of 
Pr$_2$CuO$_4$.  The resistivity is changing over nearly six orders of magnitude 
over the temperature interval.  Inset: The resistivity vs $1/T$.  This plot 
shows that at high temperatures the resistivity is activated and follows the 
form $\rho_{dc}=\rho_0 \exp(E_a/k_BT)$; a linear regression yields a value for 
$E_a=1380\pm 50$~cm$^{-1}$ ($\approx 0.17$~eV) (dotted line). 
Note the deviation from activated behavior at low temperature.}%
\label{resis} 
\end{figure}

In this paper we report on the detailed optical properties of Pr$_2$CuO$_4$ at
a variety of temperatures. The conductivity at low-frequency has a strong
temperature dependence above room temperature, which when taken with optical
estimates of the gap suggests that the transport is due to variable-range
hopping due to localized states in the gap.  Transport measurements support
this conclusion.  A strong new vibration appears close to the low-frequency
$E_u$ mode below about 200~K, in addition to other weak vibrational structure.
A normal coordinate analysis of the zone-center vibrational modes indicates
that of all the $E_u$ modes, only the low-frequency mode involves a significant
Pr-Cu interaction. It is proposed that the AFM order in this material is
responsible for the lifting of the degeneracy of the low-frequency $E_u$ mode;
the absence of any interaction between the Pr and Cu atoms in the other $E_u$
modes limits this effect to just the low-frequency $E_u$ mode.

%
%
\begin{figure*}[t]
\vspace*{-2.0cm}%
\centerline{\includegraphics[width=4.8in,angle=270]{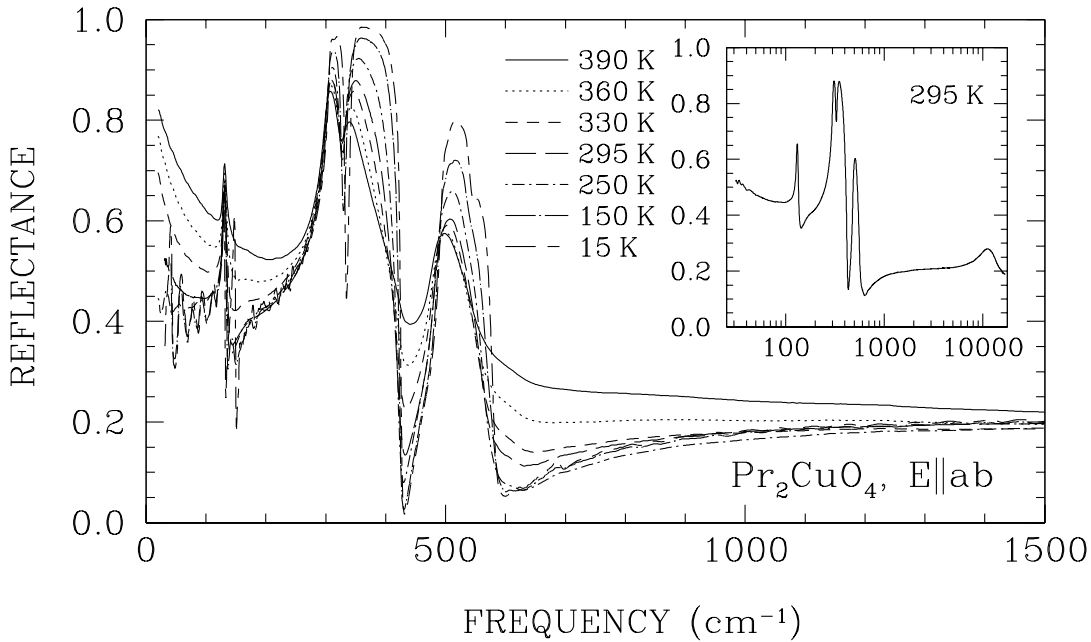}}%
\vspace*{-0.60in}%
\caption{The reflectance of Pr$_2$CuO$_4$ for light polarized in the {\it a-b} 
plane for temperatures between 15 and 390~K from $\approx 30$ to 
$1500$~cm$^{-1}$.  Above room temperature, the low frequency reflectance goes 
rapidly to unity indicating a ``metallic'' behavior, while the rapid appearance 
of the low-frequency oscillatory structure at low temperature is an indication 
of the increasingly insulating and transparent nature of the sample.  Inset: 
The reflectance at 295~K over a much wider frequency range from $\approx 30$ to 
over $16\,000$~cm$^{-1}$.}%
\label{reflec} 
\end{figure*}

%
%
\section{EXPERIMENTAL}
Single crystals of Pr$_2$CuO$_4$ were grown using a CuO-based direction
solidification technique.\cite{peng91} The crystals are thin platelets $\approx
1.5\,$mm $\times 1.5\,$mm in the {\it a-b} plane, but quite thin along the {\it
c} axis ($\approx 50\,\mu$m).  The crystals examined had a flat, mirror-like
surface which was free of flux or other residue.  Transport and optical
measurements were performed on the same sample for consistency.
%
%
The measurements of resistivity in the {\it ab}-plane were carried out using
standard four-probe configuration.  Gold wires were attached to the specimen
directly using silver-loaded epoxy and followed by a spot annealing at the
contacts with focused laser beam.  The whole specimen remained at room
temperature, except the areas of $\sim 70$~$\mu$m wide surrounding the
contacts.  This process generally resulted in a contact resistance less than
1~$\Omega$.  With such a low contact resistance, without changing the oxygen
content in these samples, we were able to measure the resistivity over six
orders of magnitude from 90~K to 340~K with high accuracy (better than 0.1\%),
shown in Fig.~\ref{resis}.  The inset shows the resistivity vs the inverse
temperature; the linear response at high temperature is an indication of
activated behavior, although this is no longer the case for $T\lesssim 160$~K.

%
%
Magnetization measurements were performed on an unoriented 2~mg crystal of
Pr$_2$CuO$_4$ in an MPMS magnetometer at an applied field of 5~T.  The behavior
of $1/\chi_M$ follows a Curie-Weiss form, and linear regression yields a
N\'{e}el temperature for Pr of $T_{N,Pr}\approx 44$~K with a Pr magnetic moment
of $3.2\,\mu_{\rm B}$.  A careful survey of the temperature region in which the
antiferromagnetic ordering of copper moments is expected $(200\,{\rm K} <
T_{N,Cu} < 320\,{\rm K})$ reveals no apparent anomaly in the magnetization. A
feature comparable in size and definition to that observed by Sun {\it et
al.}\cite{sun91} for La$_2$CuO$_4$ would have been readily observable.
However, based on the scatter in our data, an antiferromagnetic anomaly reduced
in size by more than 50\% and/or smeared over a larger temperature region might
not be discernible.  It is worth noting that the majority of studies of the
ordering of the Cu spins in these materials have been neutron scattering
measurements on substantially larger samples than the ones examined in this
work.

For the optical measurements, crystals were mounted in a cryostat on an
optically-black cone.  The temperature dependence of the reflectance was
measured at a near-normal angle of incidence from $\approx 30$ to over
$16\,000$~cm$^{-1}$ on a Bruker IFS 66v/S using an {\it in situ} overcoating
technique, which has previously been described in detail
elsewhere.\cite{homes93} This technique is especially useful when measuring
small samples, as it allows the entire face of the sample to be utilized. Above
$\approx 5000$~cm$^{-1}$ the reflectance is assumed to be temperature
independent.

%
%
\begin{figure*}[t]
\vspace*{-2.0cm}%
\centerline{\includegraphics[width=4.8in,angle=270]{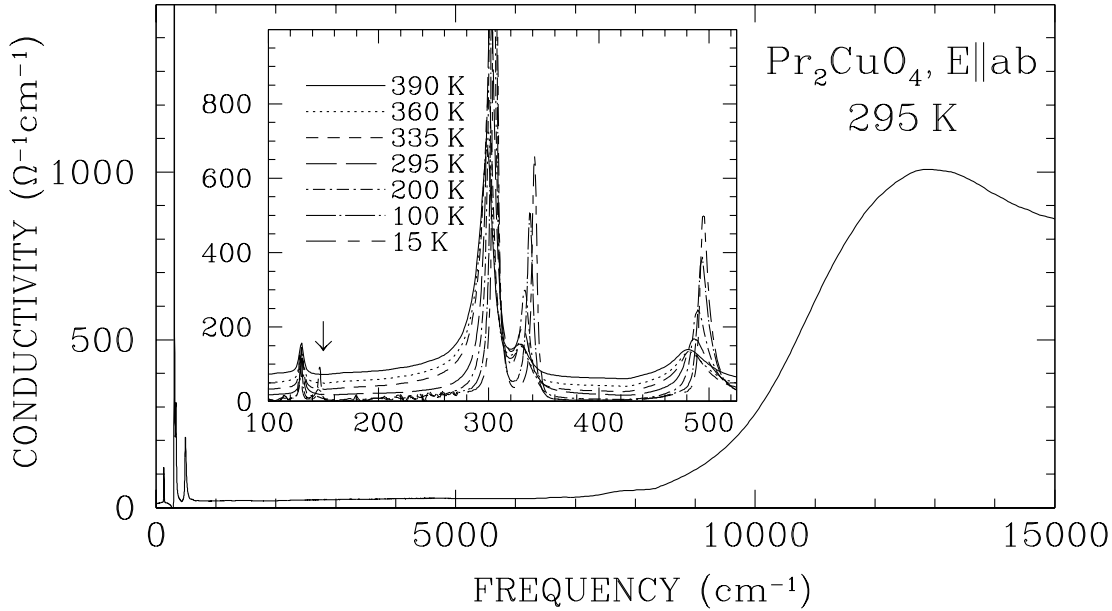}}%
\vspace*{-0.65in}%
\caption{The optical conductivity of Pr$_2$CuO$_4$ from $\approx 30$ to 
$16\,000$~cm$^{-1}$ at 295~K for light polarized in the {\it a-b} plane. Inset: 
the temperature dependence of the conductivity in the region of the four 
infrared-active $E_u$ modes; while the conductivity is dominated by the lattice 
vibrations, a broad incoherent background is observed to form rapidly above 
room temperature.  Note also the appearance of a new low-frequency vibrational 
feature at low temperature at $\approx 145$~cm$^{-1}$ (arrow). }%
\label{conduc} 
\end{figure*}

%
%
The optical conductivity has been determined from a Kramers-Kronig analysis of
the reflectance,  for which extrapolations for $\omega \rightarrow 0, \infty$
must be supplied.  At low frequency, a metallic extrapolation was used for
$T\gtrsim 260$~K, $R \propto 1 - \sqrt{\omega}$; while below this temperature
the reflectance was assumed to continue smoothly to $\approx 0.45$ at zero
frequency.  At high frequency, the reflectance was assumed to be constant above
the highest measured frequency point to $2\times 10^5$~cm$^{-1}$, above which a
free-electron ($R\propto w^{-4}$) behavior was assumed.
At low temperature, the semitransparent nature of the sample has implications
for the Kramers-Kronig analysis, which assumes specular reflectance from a
single surface only.  While the absorption due to the lattice modes assures
that the reflectance in those regions is essentially that of the bulk material,
the same cannot be said for the high-frequency region.  The presence of
multiple reflections leads to asymmetries in the line shapes and an optical
conductivity less than zero. As a result, for the calculation of the
low-frequency conductivity in the region of the lattice modes, the reflectance
has been truncated at $\approx 3000$~cm$^{-1}$ and assumed to be constant only
to $8000$~cm$^{-1}$, above which a free electron approximation has been
assumed. While the lineshapes in the conductivity are symmetric Lorentzians
whose positions and widths do not vary greatly with different choices for the
high-frequency extrapolations, the amplitudes are somewhat sensitive upon this
choice.

%
%
\section{RESULTS}
The reflectance of Pr$_2$CuO$_4$ for light polarized in the {\it a-b} plane is
shown in Fig.~\ref{reflec} from $\approx 20$ to 1500~cm$^{-1}$ at a variety of
temperatures, and in the inset from $\approx 30$ to $16\,000$~cm$^{-1}$ at room
temperature.  The reflectance at low frequency is dominated by structure due to
the normally-active infrared modes, while the reflectance at high frequency is
relatively featureless, except for some structure at $\approx 11\,000$
cm$^{-1}$. The low-frequency vibration at $\approx 130$~cm$^{-1}$ is observed
to be a single mode at room temperature. However, at low temperature this mode
is resolved as a doublet.  The low-frequency reflectance has an interesting
behavior; at low temperature the fringes indicate a lack of absorption due to
the insulating nature of the sample, but at room temperature and above the
fringes have vanished and the reflectance appears to be tending towards unity
as $\omega\rightarrow 0$; at 390~K the low frequency reflectance is over 80\%
and the system appears to be weakly metallic.

%
%
\begin{figure}[t]
\centerline{\includegraphics[width=4.0in]{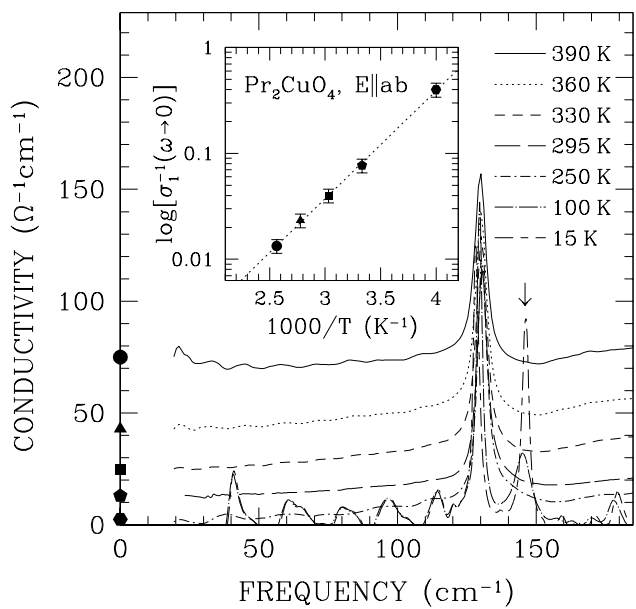}}%
\vspace*{-0.5in}%
\caption{The optical conductivity of Pr$_2$CuO$_4$ at low frequency for light 
polarized in the {\it a-b} plane from 15 to 390~K showing the rapid formation 
of the background conductivity, and the extrapolated values for $\rho_{dc} = 
1/\sigma_1(\omega\rightarrow 0)$; the different symbols denote the extrapolated 
values at different temperatures.  At low temperature a strong new vibrational 
feature is observed (arrow).  Inset: the extrapolated values for $\rho_{dc}$ vs 
$1/T$; the linear regression of the points (dotted line) yields 
$E_a=1630\pm 200$~cm$^{-1}$ ($\approx 0.2$~eV).}%
\label{lowfreq} 
\end{figure}

%
%
The optical conductivity $\sigma_1(\omega)$ calculated from a Kramers-Kronig
analysis of the reflectance at 295~K is shown in Fig.~\ref{conduc}. The low
frequency conductivity is dominated by the infrared-active $E_u$ lattice modes,
but a careful examination shows a slight asymmetry in the lineshape of the
strongest mode.  The weak feature in observed in the reflectance at $\approx
11\,000$~cm$^{-1}$ signals the onset of absorption in the conductivity at about
$9000$~cm$^{-1}$, which peaks at about $12\,000$~cm$^{-1}$.  The optical
conductivity in the region of the infrared active modes calculated using the
truncated reflectance is shown in the inset of Fig.~\ref{reflec}.  The modes
have symmetric profiles, and the three high-frequency modes all harden and
narrow with decreasing frequency.

%
%
The low frequency conductivity is shown in detail in Fig.~\ref{lowfreq} from
$\approx 20$ to 180~cm$^{-1}$.  At 390~K, there is a considerable amount of
background conductivity ($\approx 75$~$\Omega^{-1}$cm$^{-1}$) over most of the
observed frequency range, which decreases rapidly with decreasing temperature.
The inset in Fig.~\ref{lowfreq} shows the extrapolated value of the resistivity
$\rho_{dc}\approx 1/\sigma_1(\omega\rightarrow 0)$ vs $1/T$ ($T\gtrsim 250$~K);
the linear behavior indicates that the conductivity is strongly
activated.\cite{finite}

%
%
The behavior of the low-frequency mode is seen clearly in a more detailed plot
of the {\it ab}-plane reflectance shown in Fig.~\ref{detail} and the
conductivity in upper inset. The low-frequency mode can now clearly be
distinguished as fundamental at $\approx 130$~cm$^{-1}$.  A new feature appears
quickly below room temperature, gaining oscillator strength monotonically with
decreasing temperature and hardening to $\approx 145$~cm$^{-1}$ at low
temperature.

The slightly transparent nature of the sample requires that the reflections
from the back of the crystal be considered.  This approach is described in
Appendix A, and the results of fits to the reflectance using this method are
listed in Table~I.

%
%
\begin{table*}
\caption{The results of fitting the phonon parameters to the 
  reflectance of Pr$_2$CuO$_4$ for light polarized in the {\it a-b}
  plane at 295, 200 and 15~K, using the model for a lamellar 
  plate.\cite{back} A thickness of $d=50\,\mu{\rm m}$ and 
  $\epsilon_\infty = 6.5$ have been assumed.  The parameters 
  $\omega_{TO,i}$, $\gamma_i$ and $\omega_{p,i}$ refer to the 
  frequency, width and effective plasma frequency of the $i$th  
  vibration.  [All units are in cm$^{-1}$, except for the 
  dimensionless oscillator strength $S_i=\omega_{p,i}^2/
  \omega_{TO,i}^2$.] }
\begin{ruledtabular}
\begin{tabular}{cccc c cccc c cccc}
 \multicolumn{4}{c}{295~K} & & \multicolumn{4}{c}{200~K} & &
 \multicolumn{4}{c}{15~K} \\
 $\omega_{TO,i}$ & $\gamma_i$ & $\omega_{p,i}$ & $(S_i)$ &  &
 $\omega_{TO,i}$ & $\gamma_i$ & $\omega_{p,i}$ & $(S_i)$ &  &
 $\omega_{TO,i}$ & $\gamma_i$ & $\omega_{p,i}$ & $(S_i)$ \\
 \cline{1-4} \cline{6-9} \cline{11-14}
%
%
 130.6 & 4.5 & 161 & (1.51) &
 $\left\{ \begin{array}{c}  \\  \end{array} \right.$ \hspace*{-1.2cm}&
 \begin{tabular}{c}  130.4  \\ 141.1   \end{tabular} &
 \begin{tabular}{c}    3.3  \\   9.9   \end{tabular} &
 \begin{tabular}{c}  157    \\  69     \end{tabular} &
  \begin{tabular}{c} (1.45) \\ (0.24) \end{tabular} &
 $\left\{ \begin{array}{c}  \\  \end{array} \right.$ \hspace*{-1.2cm} &
 \begin{tabular}{c}  128.2  \\ 146.1   \end{tabular} &
 \begin{tabular}{c}    2.7  \\   2.4   \end{tabular} &
 \begin{tabular}{c}  144    \\ 124     \end{tabular} &
  \begin{tabular}{c} (1.26) \\ (0.72)  \end{tabular} \\
%
%
  304 & 7.5 & 815 & (7.18) & \ &
  305 & 3.0 & 831 & (7.42) & \ &
  306 & 1.4 & 834 & (7.43) \\
%
%
  331 & 17.4  & 469 & (2.01) &  &
  333 &  7.7  & 458 & (1.89) &  &
  341 &  3.4  & 431 & (1.59) \\
%
%
  490 & 27.3  & 511 & (1.08) &  &
  491 & 19.5  & 508 & (1.07) &  &
  495 &  8.4  & 503 & (1.03) \\
%
%
       &      &     &        &  &
   541 &   38 & 112 & (0.04) &  &
   542 &   25 & 108 & (0.04) \\
%
%
       &      &     &        &  &
   688 &   25 &  60 & (0.01) &  &
   688 &   15 &  90 & (0.02) \\
\end{tabular}
\end{ruledtabular}
%
%
\label{phonons} 
\end{table*}
%

%
%
\section{DISCUSSION}
%
%
\subsection{Electronic properties}
The peak in the optical conductivity of Pr$_2$CuO$_4$ at $\approx
12\,500$~cm$^{-1}$ in Fig.~\ref{conduc} has been observed in previous
work.\cite{lupi92,cooper90,arima93} The location of this feature is similar to
the onset of absorption in La$_2$CuO$_4$ and related materials\cite{perkins98}
at $\approx 11\,000$~cm$^{-1}$ ($\approx 1.4$~eV). The peak is characteristic
of a semiconducting band edge or a polaronic excitation.  In either of these
cases, the {\it dc} resistivity is expected to be activated and to follow the
form\cite{alexandrov}
\begin{equation}
  \rho_{dc}\propto e^{E_a/{k_BT}},
  \label{active}
\end{equation}
where $E_a$ is either the half the optical gap $2\Delta$, or half the polaronic
level shift $E_p$. The low-frequency reflectance increases quickly above room
temperature, which is indicative of a ``metallic'' response in which the
reflectance goes to unity at zero frequency. This observation is realized in
the optical conductivity, where $\sigma_{dc} \equiv \sigma_1(\omega \rightarrow
0)$ has a very strong temperature dependence for $T \gtrsim 300$~K. The plot of
$\rho_{dc}=1/\sigma_{dc}$ vs $1/T$ is shown in the inset in Fig.~\ref{lowfreq},
and is described quite well by Eq.~(\ref{active}); a linear regression yields
$E_a\approx 1630\pm 200$~cm$^{-1}$ ($\approx 0.2$~eV).  The rather large error
associated with this estimate is due to the fact that the fit is limited to a
narrow interval in the high-temperature region.  The transport data shown in
Fig.~\ref{resis} and the inset are also described quite well by activated
behavior; a linear regression applied in the high-temperature region ($T\gtrsim
160$~K) yields $E_a=1380\pm 50$~cm$^{-1}$ ($\approx 0.17$~eV) and is shown as a
dashed line in the inset of Fig.~\ref{resis}; this value for $E_a$ is fairly
close to the value determined from the optical conductivity.
%
%
The optical conductivity of the nickelates, which have a similar appearance,
has been fitted using a small-polaron model,\cite{eklund93, katsufuji96,
alexandrov} which for $T\gg \omega/2$ produces an asymmetric Gaussian peak with
a maximum at $4E_a$.  This model would imply that the peak in the conductivity
at high temperature should occur at $\approx 6000$~cm$^{-1}$, which is much
less than the observed value of $\approx 12\,500$~cm$^{-1}$.  A rough estimate
of the direct optical gap may be made by extrapolating from the linear part of
the leading edge conductivity in Fig.~\ref{conduc} to the abscissa, which gives
$2\Delta\approx 9500$~cm$^{-1}$. If the transport were due to the carrier pair
density from thermal excitations across the gap, then $\sigma_{dc} \propto
n_i\approx \exp(-\Delta/k_BT)$. However, $E_a \ll \Delta$, suggesting that the
{\it dc} transport is due to variable range hopping between localized states
within the gap.\cite{mott}
%
%
\begin{figure}
\centerline{\includegraphics[width=4.0in]{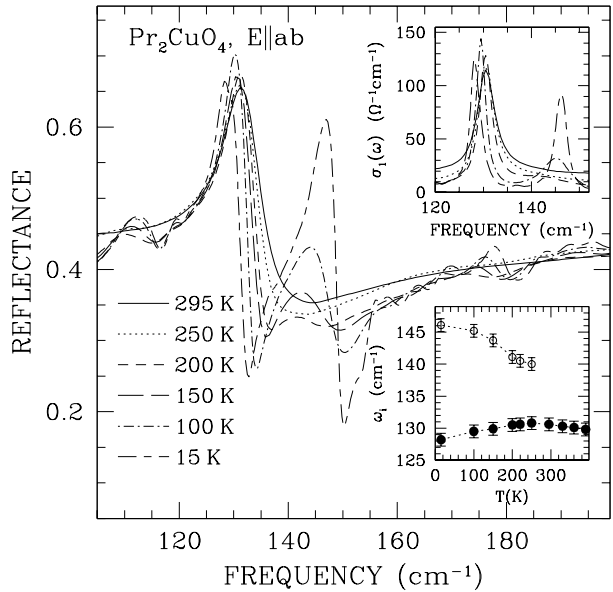}}%
\vspace*{-0.5in}%
\caption{The reflectance of Pr$_2$CuO$_4$ for light polarized in the {\it a-b} 
planes in the region of the low frequency infrared-active $E_u$ mode.  The 
vibration is a single mode at room temperature, but is resolved as a doublet at 
low temperature.  Upper inset: The optical conductivity which shows not only 
the appearance of the new mode, but also the rapid increase in strength and 
narrowing of this feature. Lower inset: The temperature dependence of the 
fundamental mode (filled circles) and the new vibrational feature (open 
circles).}%
\label{detail} 
\end{figure}
%
%
At low temperatures, the resistivity due to hopping is not activated, but
instead has the form
\begin{equation}
  \rho_{dc}\propto e^{(T_0/T)^{1/n}},
\end{equation}
where $T_0$ is a characteristic temperature, and $n=d+1$ and $d$ is the
dimensionality of the system.\cite{mott} The transport measurements have been
performed over a wide enough range so that the departure from activated to
power-law behavior at low temperature is clearly visible in the insert in
Fig.~\ref{resis}. Different power laws were examined, but it appears that the
resistivity at low temperature is well described by $\rho_{dc}\propto
\exp{[(T_0/T)^{1/4}]}$, suggesting that the hopping is a three-dimensional
rather than a two-dimensional phenomena restricted to the CuO$_2$
planes.\cite{ambegaokar71} The localized states that result in hopping may
arise from defects which produced states within the gap.  There is a very weak
feature in the optical conductivity in Fig.~\ref{conduc} at $\approx
8000$~cm$^{-1}$ which occurs at $\approx E_a$ below the leading edge of the
absorption.  This feature may suggest a possible origin for the localized
states.  However, it not clear if the defects states responsible for this
feature are an intrinsic property of the sample, or due to extrinsic effects.

%
%
\subsection{Vibrational properties}
\subsubsection{Normal coordinate analysis of the lattice modes}
The unusual behavior of the low-frequency $E_u$ mode requires a detailed
understanding not only of the nature of this particular mode, but of the other
$E_u$ modes as well.  For this reason, a normal coordinate analysis of
Pr$_2$CuO$_4$ was undertaken.\cite{normal}

The normal coordinate calculations were performed using Wilson's $GF$ matrix
method \cite{wilson,shimanouchi61} and a commercially available software
package.\cite{vibratz}  Initially, a simple valence force field was adopted,
consisting of bond stretching and angle bending coordinates; the types and
values of the force constants used, and the internal coordinates to which they
correspond are listed in Table~II.  The force field involves six bond
stretching force constants and four angle bending force constants.  The bond
stretch and angle bends all involve oxygen bonding, with the exception of the
bond stretches denoted $f_4$ and $f_6$, which deal with the Pr-Pr and Pr-Cu
interactions; while the value for the Pr-Cu bond stretch is relatively small,
the refinement is significantly worse if it is omitted.  While this initial
force field described the vibrations in the Cu-O(1) planes quite well, the
corrugated nature of the Pr-O(2) layers shown in Fig.~\ref{unitcell} leads to
difficulties in describing the restoring forces within this layer. The
agreement with the observed frequencies improved significantly when an
interaction force constant between the bond stretch and the angle bend in the
Pr-O(2) layer was introduced. The comparison between observed and calculated
frequencies, and the potential energy distribution (PED) are listed in
Table~III. The PED's gives the relative contribution of the force constants to
the potential energy of the normal modes.  This treatment yields the atomic
displacements and PED for all of the normal modes.  Each of the in-plane
infrared active $E_u$ modes merits further discussion.

%
%
\begin{table}
\caption{Force constants and internal coordinates for Pr$_2$CuO$_4$.  The 
labeling scheme of the atoms is shown in Fig.~1.} 
\begin{ruledtabular}
\begin{tabular}{cclcc}
  Force      &   Internal      &     & Distance (\AA )/ & \\
  constant   &  coordinate     &     & Angle ($^\circ$) & Value$^a$ \ \\
  \cline{1-5}
  & & & &  \\
  \multicolumn{3}{c}{Bond stretch} & & \\
  \cline{1-3}
  $f_1$      & Cu-O(1)         & [4]$^b$ & 1.98        & 0.355 \\
  $f_2$      & Pr-O(2)         & [8] & 2.12            & 0.532 \\
  $f_3$      & Pr-O(1)         & [8] & 3.03            & 0.785 \\
  $f_4$      & Pr-Pr           & [1] & 4.58            & 0.787 \\
  $f_5$      & O(2)-O(2)       & [4] & 2.80            & 0.300 \\
  $f_6$      & Pr-Cu           & [8] & 3.62            & 0.067 \\
  & & & &  \\
  \multicolumn{3}{c}{Angle bend} & & \\
  \cline{1-3}
  $\alpha_1$ & O(1)-Cu-O(1)    & [4] &  90             & 0.572 \\
  $\alpha_2$ & O(1)-Cu-O(1)$^c$& [4] & 180             & 0.195 \\
  $\alpha_3$ & O(2)-Pr-O(2)    & [4] & 138             & 0.469 \\
  $\alpha_4$ & O(1)-Pr-O(1)    & [4] &  82             & 0.478 \\
  & & & & \\
  \multicolumn{3}{c}{Interaction} & & \\
  \cline{1-3}
  $i_1$      & $f_2\alpha_3$   & [16] &                & 0.074 \\
\end{tabular}
\end{ruledtabular}
\footnotetext[1] {Bond stretches are in units of md/\AA , angle bends in md-\AA /rad.}%
\footnotetext[2] {The number of internal coordinates.}%
\footnotetext[3] {In plane and out of plane.}%
\end{table}

%
%
%
The $E_u$ mode at 490~cm$^{-1}$ at room temperature involves primarily the
in-plane Cu-O angle bending resulting in displacements of the O(1) atoms.  In
addition, there are small displacements of the atoms in the Pr-O(2) planes; the
motions of the Pr and O(1) atoms are coordinated.  However, as Table~III shows,
there is no interaction between the rather small Pr and the Cu displacements.
In general, the eigenvectors determined from the normal coordinate analysis are
in good agreement with the calculated zone-center displacements shown in
Ref.~\onlinecite{zhang91}.  This mode hardens to 495~cm$^{-1}$ at 15~K and
narrows dramatically from $\approx 25$ to $\approx 8$~cm$^{-1}$, as shown in
Table~I. The oscillator strength of this mode does not vary with temperature.
The slight asymmetry that is observed in the optical conductivity of the
high-frequency $E_u$ mode in Fig.~\ref{conduc} is most likely produced by the
uncertainty in the high-frequency extrapolation used in the Kramers-Kronig
analysis; fits to the reflectance are exact and do not suggest a large
asymmetry in this feature or others (Appendix A).

%
%
The mode at 331~cm$^{-1}$ is a pure Pr-O(2) bond stretch, while the
304~cm$^{-1}$ mode is almost a pure Cu-O(1) bond stretch (Table~III).  The mode
at 331~cm$^{-1}$ hardens considerably to 341~cm$^{-1}$ at 15~K, and also
narrows from $\approx 17$ to 3.4~cm$^{-1}$.
%
%
The mode at $\approx 304$~cm$^{-1}$ hardens only slightly with decreasing
temperature, but narrows significantly from $\approx 8$~cm$^{-1}$ at room
temperature to 1.4~cm$^{-1}$ at 15~K; this is the strongest $E_u$ mode with an
oscillator strength of $S \approx 7.4$.

%
%
The mode at 130~cm$^{-1}$ is a combination of Cu-O angle bending, resulting in
the in-plane displacements of the Cu and O(1) atoms, as well as the
out-of-phase motion of the Pr and O(2) atoms in the Pr-O layers; interestingly,
this mode has a significant coordination between the displacements of the Pr
atoms and the Cu and O(1) atoms, indicated in Table~III. The only other mode
that has a strong Pr-Cu interaction is the low-frequency {\it c}-axis $A_{2u}$
mode. The low-frequency $E_u$ mode softens to $\approx 128$~cm$^{-1}$ at low
temperature, while narrowing. The observed frequencies are in good agreement
with previous work,\cite{tajima91,crawford90b} however, the linewidths are all
narrower and there is only rough agreement with the reported strengths.

%
%
%
\subsubsection{Origin of the low temperature doublet at $\approx 130$~cm$^{-1}$}
The most unusual feature in the reflectance spectra is striking appearance of a
new mode at low temperature, which is just above the low frequency $E_u$ mode.
The new mode is not simply a peculiar artifact due to fringes in the presence
of vibrational structure, as Fig.~\ref{model}(b) demonstrates (Appendix A).
While this feature has been previously observed at low
temperature\cite{crawford90a,crawford90b} the evolution of this new mode has
never been studied or fully explained.  The new feature appears suddenly below
$\lesssim 250$~K at $\approx 140$~cm$^{-1}$, as seen in Fig.~\ref{detail} and
in the insets; this mode hardens dramatically and narrows with decreasing
temperature, gaining strength monotonically until it has almost the same
oscillator strength as the low-frequency $E_u$ mode (Table I).  This feature
does not appear to evolve from a strong asymmetry or shoulder in the
low-frequency $E_u$ mode.
There is also additional very weak new fine structure in the low temperature
reflectance at $\approx 541$ and 688~cm$^{-1}$.  These frequencies are well
above the high-frequency $E_u$ mode at $\approx 490$~cm$^{-1}$.  A
symmetry-breaking process would allow Raman-active modes as well as the
longitudinal optic (LO) modes to become weakly active. However, the highest
observed frequency for a Raman mode is the $E_g$ mode at $\approx
480$~cm$^{-1}$, suggesting that these features are not Raman modes.  To
determine the positions of the LO modes, the reflectance and optical
conductivity have also been fit using a factorized form of the dielectric
function\cite{servoin80}
\begin{equation}
  \tilde\epsilon(\omega)=\epsilon_\infty \prod_j {
  {\omega_{LO,j}^2-\omega^2-i\gamma_{LO,j}\omega} \over
  {\omega_{TO,j}^2-\omega^2-i\gamma_{TO,j}\omega} },
\end{equation}
%
%
where $\omega_{LO,j}$, $\omega_{TO,j}$, $\gamma_{LO,j}$ and $\gamma_{TO,j}$ are
the $j$th LO and the normally-active transverse optic (TO) modes, and LO and TO
damping, respectively. The fit at 15~K yields values for the TO modes which are
nearly identical to those shown in Table~I, and LO modes at 131, 149, 332, 430
and 588 cm$^{-1}$, respectively. None of the LO modes are close to the new
features observed at low temperature, making it unlikely that any new structure
is due to the LO modes.

%
%
A more promising explanation of the splitting of the low frequency mode may
involve the fact that of all the $E_u$ modes, only the low-frequency mode
involves a significant Pr-Cu interaction.  Magnetization measurements in this
work show that the Pr atoms are ordered at low temperature ($\lesssim 50$~K),
and neutron scattering measurements\cite{sumarlin95} show that there is at
least a partial ordering at higher temperatures, indicating the presence of a
Pr-Cu exchange interaction.  In the case of the low-frequency $E_u$ mode, the
displacements of the Pr and Cu atoms are strongly correlated, suggesting that
the exchange interaction will have a significant effect upon the nature of this
vibration. This can lead to a symmetry-breaking process and the splitting of
the doubly-degenerate $E_u$ mode.  This view is consistent with the observation
that this new feature is {\it not} present in Pr$_{1.85}$Ce$_{0.15}$CuO$_4$,
which is in the region of the phase diagram for this class of materials where
the AFM order is destroyed.\cite{heyen90}  Because the other $E_u$ modes do not
involve strong Pr-Cu coupling, no exchange-induced splitting would be expected.
In fact, of all the other vibrational modes, only the low-frequency $A_{2u}$
mode has a significant Pr-Cu interaction (Table~III), but since this mode is
singly degenerate no splitting is expected, nor it is
observed.\cite{crawford90b}
%
%
Similar behavior might also be expected in Nd$_2$CuO$_4$, where the Nd moments
order below 37~K.  In fact, a weak feature is observed in the reflectance of
Nd$_2$CuO$_4$ at 10~K at $\approx 165$~cm$^{-1}$, quite close to the
low-frequency $E_u$ mode, and it has been suggested that this feature is indeed
magnetic in origin.\cite{heyen90} Kramers doublets have also been reported in
the Raman spectra at low temperature,\cite{dufour95} which have been attributed
to the removal of degeneracy by the Nd-Cu exchange interaction.\cite{jandl93}
While this is a reasonable explanation for the strong feature at low frequency,
it is less satisfactory for the fine structure observed at high frequency.

%
%
\begin{table}
\caption{Calculated and observed frequencies for the zone-center vibrations of 
Pr$_2$CuO$_4$ at room temperature, and the potential energy distribution (PED). 
(All frequencies are in cm$^{-1}$.) } 
\begin{ruledtabular}
\begin{tabular}{cccl}
  Symmetry   & obs.$^a$ & cal. & PED$^b$ ($\geq 10$\% ) \\
\hline 
  $A_{1g}$ & 228      & 227 & $f_3(42), f_4(37), \alpha_3(17)$ \\
  $B_{1g}$ & 328      & 326 & $f_2(28), \alpha_3(73)$ \\
  $E_g$    & 480      & 481 & $f_2(46), f_5(54)$ \\
  $E_g$    & 126      & 123 & $f_2(18), f_3(52), f_5(23)$ \\
\hline 
  $A_{2u}$ & 505      & 504 & $f_3(83)$ \\
  $A_{2u}$ & 271      & 271 & $f_2(43), \alpha_3(113), i_1(-58)$ \\
  $A_{2u}$ & 135      & 135 & $f_6(37), \alpha_2(58)$ \\
  $B_{2u}$ & (silent) & 480 & $f_3(82)$ \\
  $E_u$    & 490      & 490 & $f_3(60), \alpha_1(29), \alpha_2(10)$ \\
  $E_u$    & 331      & 332 & $f_2(98)$ \\
  $E_u$    & 304      & 303 & $f_1(93)$ \\
  $E_u$    & 131      & 129 & $f_3(28), f_6(22), \alpha_1(30), \alpha_2(10)$ \\
\end{tabular}
\end{ruledtabular}
\footnotetext[1] {The observed frequencies for the $A_{2u}$ modes are taken 
  from Ref.~\onlinecite{tajima91}, while the Raman active modes are taken as
  those for Nd$_2$CuO$_4$ from Ref.~\onlinecite{jandl93}.  The frequencies for
  the $E_u$ modes are from this work (Table I).}%
\footnotetext[2] {See Table~II for the identification of the force constants.} 
\end{table}

%
%
The new structure observed at $\approx 688$~cm$^{-1}$ at low temperature is
similar in frequency to structure that has been observed the reflectance in
other electron-doped materials\cite{calvani96,wang90} at $\approx
690$~cm$^{-1}$.  It has been suggested that much of the fine structure observed
in the cuprate systems are local vibrational modes due to polaron fine
structure.\cite{calvani96,lupi98} Alternatively, a number of weak features are
also seen below the N\'{e}el transition in CuO, including a mode at $\approx
690$~cm$^{-1}$, which has been attributed to the activation of a zone-boundary
mode due to the reduction of the Brillouin zone resulting from the magnetic
order.\cite{chen95,kuzmenko01}
It is possible that the magnetic order is responsible for a weak structural
distortion and the activation of zone boundary modes due to the commensurate
reduction of the Brillouin zone.  However, a difficulty with this explanation
is that any reduction of the Brillouin zone should result in a many new modes
being activated, rather than just the three that are observed.
One of the difficulties in explaining the vibrational fine structure observed
in this system is the failure to observe these results on a consistent basis in
different studies.\cite{calvani96,homes97}  This raises the distinct
possibility that some of these features arise from impurities and/or phase
separation, ultimately making it difficult to identify a specific mechanism
responsible for the origin of the fine structure.
%
%

%
%
\section{Conclusions}
The strong temperature dependence of the low-frequency reflectance above
$\approx 250$~K results in a resistivity $\rho_{dc}\approx 1/
\sigma_1(\omega\rightarrow 0)$ which is exponential with temperature, following
the functional form $\rho_{dc} \propto \exp(E_a/k_BT)$, where $E_a\approx
0.20$~eV; a more detailed transport measurement yields a slightly lower and
more accurate value of $E_a\approx 0.17$~eV.  The fact that the transport gap
$E_a$ is much less than the inferred optical gap $2\Delta\approx 1.2$~eV, and
that transport measurements of the resistivity show a departure from activated
behavior for $T\lesssim 160$~K to a $\rho_{dc} \propto \exp[(T_0/T)^{1/4}]$
power-law behavior suggests that the transport in this material is due to
variable-range hopping between localized states in the gap.

Due to the unusual vibrational structure in Pr$_2$CuO$_4$ at low temperature,
considerable attention was devoted to a discussion of the lattice modes.  A
normal coordinate analysis has been performed to provide a detailed
understanding of the nature of the zone-center vibrations in this material. A
prominent feature close to the low-frequency $E_u$ mode develops at low
temperature, and additional weak fine structure is observed at high frequency.
These new features appear below $T_{N,Cu}$ at roughly the same temperature at
which the Pr moments are thought to begin to order ($\approx 200$~K).
Only the low-frequency $E_u$ mode involves a significant Pr-Cu interaction.
This suggests that the new vibrational structure is due to the
antiferromagnetic order in this material and the removal of degeneracy of this
mode.

%
%
\begin{acknowledgments}
We are grateful to A.R.~Moodenbaugh for performing the magnetization
measurements.  We would also like to thank D.N.~Basov, V.J.~Emery,
J.L.~Musfeldt, M.~Strongin, T.~Timusk, J.M.~Tranquada, and J.J.~Tu for many
helpful discussions.
This work was supported by the Department of Energy under contract number
DE-AC02-98CH10886; work in Maryland is supported by the NSF Condensed Matter
Physics Division under grant No.~DMR 9732736.  Research undertaken at NSLS was
supported by the U.S. DOE, Division of Materials and Chemical Sciences.
\end{acknowledgments}

%
%
\appendix
\section{Reflectance of a dielectric slab}
%
%
Due to the complications introduced by the slightly transparent nature of the
sample, the reflectance and the lattice modes have been fit using a model which
considers reflections from both the front and back surfaces of the crystal. The
frequency-dependent reflectance of a lamellar plate at a normal angle of
incidence is $R=\tilde{r}_l \tilde{r}_l^\ast$, where $\tilde{r}_l$ is defined
as
%
%
\begin{figure}
\centerline{\includegraphics[width=4.0in]{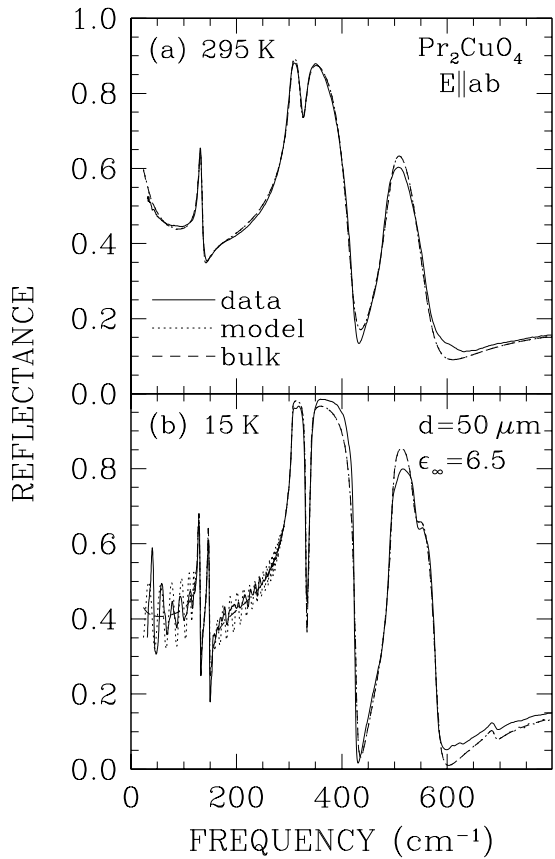}}%
\vspace*{-0.5in}%
\caption{(a) The reflectance in the infrared region of Pr$_2$CuO$_4$ for light 
polarized in the {\it a-b} plane at room temperature (solid line) and the fit 
to the data from Eq.~(1) (dotted line) using the vibrational parameters in 
Table~I, with $\epsilon_\infty = 6.5$ and a sample thickness of $d=50$~$\mu$m. 
The dashed line indicates the calculated behavior the (infinitely thick) bulk 
material, which is identical to that of the model, indicating the opaque nature 
of the sample. 
(b) The reflectance at 15~K (solid line) with the model fit (dotted line) using 
the vibrational parameters in Table~I, using the same values for $d$ and 
$\epsilon_\infty$.  The dashed line indicates the calculated behavior of the 
bulk material.  The doublet structure, fringe spacing and amplitude are in good 
agreement with the reflectance.  Note the absence of fringes in regions of 
strong absorption.}%
\label{model} 
\end{figure}
\begin{equation}
  \tilde{r}_l={ {\tilde r \left[ 1-e^{i\omega 4\pi\tilde{n}d}\right]} \over
  {1-r^2e^{i\omega 4\pi\tilde{n}d}} },
\end{equation}
where $d$ is the sample thickness, and $\tilde{r}$ is the Fresnel reflectance
of the bulk material,
\begin{equation}
  \tilde{r}={ {1-\tilde{n}} \over {1+\tilde{n}} }.
\end{equation}

The complex refractive index is $\tilde{n}=n+ik$, which is related to the
complex dielectric function $\tilde\epsilon = \epsilon_1 + i\epsilon_2 =
\tilde{n}^2$, allowing the real and imaginary parts of the refractive index $n$
and $k$ to be determined
\begin{equation}
 n= \left\{ {1\over 2} \left[ \sqrt{ \epsilon_1^2 + \epsilon_2^2 } +
    \epsilon_1 \right] \right\}^{1/2},
\end{equation}
\begin{equation}
 k= \left\{ {1\over 2} \left[ \sqrt{ \epsilon_1^2 + \epsilon_2^2 } -
    \epsilon_1 \right] \right\}^{1/2}.
\end{equation}
The optical properties of the bulk material are modeled using a series of
Lorentzian oscillators
\begin{equation}
  \tilde\epsilon(\omega)=\epsilon_\infty+\sum_j {{\omega_{p,j}^2} \over
  {\omega_j^2-\omega^2-i\gamma_j\omega}},
  \label{lorentz}
\end{equation}
where $w_j$, $\gamma_j$ and $\omega_{p,j}$ are the frequency, width and
effective plasma frequency of the $j$th vibration, and $\epsilon_\infty$ is the
core contribution to the dielectric function. The dimensionless oscillator
strength is written as $S_j = \omega_{p,j}^2/ \omega_j^2$.  The optical
conductivity $\tilde\sigma(\omega) = \sigma_1(\omega) + i\sigma_2(\omega)$ is
related to the complex dielectric function by $\tilde\sigma(\omega) = -i\omega
\tilde\epsilon(\omega) /4\pi$.

The phonon parameters were refined by a non-linear least-squares fit of the
model for the reflectance of a thin dielectric slab to the experimental
reflectance in Fig.~\ref{reflec}.  The results at 295, 200 and 15~K are listed
in Table~I, and a comparison of the experimental reflectance and the model
results are shown in Figs.~\ref{model}(a) and \ref{model}(b) at 295 and 15~K,
respectively.  The broad, incoherent electronic background observed at room
temperature is modeled by a zero-frequency Drude term, as well as two
mid-infrared overdamped oscillators (Table~I).  The results were then compared
with fits to the features in the conductivity by assuming simple linear
background in the region of the lattice modes, and were found to be in good
agreement. This indicates that in the region of the phonon features where the
sample is opaque, the Kramers-Kronig relation yields acceptable values for the
conductivity.

%
%
%
\bibliography{pcoref}

\begin{thebibliography}{56}
\bibitem[{fin()}]{finite}
\bibinfo{note}{This finite conductivity effectively renders the sample opaque
  at room temperature and above. The origin of the fringes at low temperature
  is attributed to interference effects due to reflections from the back of the
  crystal as the sample becomes increasingly transparent, as evidenced by the
  unphysical results $[\sigma_1(\omega) < 0]$ at low temperature.}
\bibitem[{nor()}]{normal}
\bibinfo{note}{Normal coordinate calculations, which are applicable to zero
  wavevector normal mode analysis, have an advantage over lattice dynamical
  calculations in that non-central forces such as those involved in angle
  bending may be treated. The effect of angle-bending force constants is
  proportional to the covalency of the compound (Ref.~\onlinecite{martin70});
  given the estimated covalency of 0.5 for the cuprate systems (Ref.
  \onlinecite{pauling88}) they should be considered.}
\end{thebibliography}

\end{document}